\documentclass{nature2}

\usepackage{graphicx}        % standard LaTeX graphics tool
\usepackage[utf8]{inputenc}
\usepackage{hyperref}
\usepackage{wasysym}
\usepackage{bookmark}
\usepackage{longtable}
\usepackage{bm}
\usepackage[british]{babel}
\usepackage{color}

\usepackage{bibunits}
\usepackage{etoolbox}
\usepackage{lipsum}

\defaultbibliographystyle{naturemag}

\definecolor{orange}{RGB}{255,127,0}
\definecolor{darkgreen}{RGB}{8,114,11}

\makeatletter
\newcommand{\ion}{}% Check, that \ion is not yet defined
\DeclareRobustCommand*{\ion}[2]{%
  #1$\;$%
  \if b\expandafter\@car\f@series\relax\@nil
    \begingroup % keep changes to scratch box register 0 local
      \sbox0{\rmfamily\mdseries\textsc{v}}%
      \resizebox{!}{\ht0}{\rmfamily\@Roman{#2}}%
    \endgroup
  \else
    \textsc{\rmfamily\@roman{#2}}%
  \fi
}
\makeatother

\makeatletter
\newcounter{mybibcounter}
\renewenvironment{thebibliography}[1]
     {\section*{\refname}%
      \@mkboth{\MakeUppercase\refname}{\MakeUppercase\refname}%
      \list{\@biblabel{\@arabic\c@mybibcounter}}%
           {\settowidth\labelwidth{\@biblabel{#1}}%
            \leftmargin\labelwidth
            \advance\leftmargin\labelsep
            \@openbib@code
            \@nmbrlisttrue\def\@listctr{mybibcounter}%
            \let\p@mybibcounter\@empty
            \renewcommand\themybibcounter{\@arabic\c@mybibcounter}}%
      \sloppy
      \clubpenalty4000
      \@clubpenalty \clubpenalty
      \widowpenalty4000%
      \sfcode`\.\@m}
     {\def\@noitemerr
       {\@latex@warning{Empty `thebibliography' environment}}%
      \endlist}
\makeatother

\def\kms{km.s$^{-1}$}         %m.s -1
\def\ms{\hbox{m.s$^{-1}$}}         %m.s -1
\def\cmss{\hbox{cm.s$^{-2}$}}       %cm.s -2
\def\gcm3{\hbox{g.cm$^{-3}$}}       %g.cm-3
\def\vsini{\hbox{$\upsilon \sin i_{\star}$}}      %vsini
\def\Msun{\hbox{M$_{\astrosun}$}}             %Msun
\def\Rsun{\hbox{R$_{\astrosun}$}}

\def\Mearth{\hbox{M$_{\oplus}$}}
\def\Rearth{\hbox{R$_{\oplus}$}}
\def\degr{\hbox{$^\circ$}}
\def\teff{T$_{\rm eff}$}
\def\logg{log~{\it g}}
\def\met{[Fe/H]}

\title{An Earth-sized exoplanet with a Mercury-like composition}

\author{A. Santerne$^{\ref{LAM}}$,  %OK
           B. Brugger$^{\ref{LAM}}$,    %OK
           D. J. Armstrong$^{\ref{warwick}}$,     %OK
           V. Adibekyan$^{\ref{IA}}$,    %OK
           J. Lillo-Box$^{\ref{ESO}}$,   %OK
           H. Gosselin$^{\ref{LAM}, \ref{UPS}}$,  %OK
           A. Aguichine$^{\ref{LAM},\ref{ENS}}$,  %OK
           J.-M. Almenara$^{\ref{geneva}}$,
           D. Barrado$^{\ref{CAB}}$,         %OK
           S. C. C. Barros$^{\ref{IA}}$,    %OK
           D. Bayliss$^{\ref{geneva}}$,      %OK
           I. Boisse$^{\ref{LAM}}$,             %OK
           A. S. Bonomo$^{\ref{Torino}}$,   %OK
           F. Bouchy$^{\ref{geneva}}$,       %OK
           D. J. A. Brown$^{\ref{warwick}}$,   %OK
           M. Deleuil$^{\ref{LAM}}$,            %OK
           E. Delgado~Mena$^{\ref{IA}}$,   %OK
           O. Demangeon$^{\ref{IA}}$,      %OK
           R. F. D\'iaz$^{\ref{UBA},\ref{CONICET},\ref{geneva}}$,   %OK
           A. Doyle$^{\ref{warwick}}$,      %OK
           X. Dumusque$^{\ref{geneva}}$,     %OK
           F. Faedi$^{\ref{warwick},\ref{catania}}$,             %OK
           J. P. Faria$^{\ref{IA},\ref{UPorto}}$,
           P. Figueira$^{\ref{ESO},\ref{IA}}$,     % OK
           E. Foxell$^{\ref{warwick}}$,     % OK
           H. Giles$^{\ref{geneva}}$,
           G. H\'ebrard$^{\ref{IAP},\ref{OHP}}$,
           S. Hojjatpanah$^{\ref{IA},\ref{UPorto}}$,          
           M. Hobson$^{\ref{LAM}}$,       %OK
           J. Jackman$^{\ref{warwick}}$,  %OK
           G. King$^{\ref{warwick}}$,     %OK
           J. Kirk$^{\ref{warwick}}$,     % OK
           K. W. F. Lam$^{\ref{warwick}}$,     %OK
           R. Ligi$^{\ref{LAM}}$,        % OK
           C. Lovis$^{\ref{geneva}}$,      %OK
           T. Louden$^{\ref{warwick}}$,     %OK
           J. McCormac$^{\ref{warwick}}$,
           O. Mousis$^{\ref{LAM}}$,           %OK
           J. J. Neal$^{\ref{IA},\ref{UPorto}}$,    %OK
           H. P. Osborn$^{\ref{warwick},\ref{LAM}}$,       %OK
           F. Pepe$^{\ref{geneva}}$,
           D. Pollacco$^{\ref{warwick}}$,     %OK
           N. C. Santos$^{\ref{IA},\ref{UPorto}}$,     %OK
           S. G. Sousa$^{\ref{IA}}$,     %OK
           S. Udry$^{\ref{geneva}}$      %OK
           \& A. Vigan$^{\ref{LAM}}$}    %OK

\begin{document}

\maketitle

\begin{affiliations}
 \item Aix Marseille Univ, CNRS, CNES, LAM, Marseille, France\label{LAM}
  \item Department of Physics, University of Warwick, Gibbet Hill Road, Coventry, CV4 7AL, UK\label{warwick}
  \item Instituto de Astrof\'isica e Ci\^{e}ncias do Espa\c co, Universidade do Porto, CAUP, Rua das Estrelas, 4150-762 Porto, Portugal\label{IA}
 \item European Southern Observatory (ESO), Alonso de Cordova 3107, Vitacura, Casilla 19001, Santiago de Chile, Chile\label{ESO}
 \item  Universit\'e de Toulouse, UPS-OMP, IRAP, Toulouse, France\label{UPS}
 \item Paris-Saclay Universit\'e, ENS Cachan, 61 av. du Pr\'esident Wilson, 94230 Cachan, France\label{ENS}
 \item Observatoire Astronomique de l'Universit\'e de Gen\`eve, 51 Chemin des Maillettes, 1290 Versoix, Switzerland\label{geneva}
 \item Depto. de Astrof\'isica, Centro de Astrobiolog\'ia (CSIC-INTA), ESAC campus 28692 Villanueva de la Ca\~nada (Madrid), Spain\label{CAB}
 \item INAF -- Osservatorio Astrofisico di Torino, Strada Osservatorio 20, I-10025, Pino Torinese (TO), Italy\label{Torino}
 \item Universidad de Buenos Aires, Facultad de Ciencias Exactas y Naturales. Buenos Aires, Argentina\label{UBA}
 \item CONICET - Universidad de Buenos Aires. Instituto de Astronom\'ia y F\'isica del Espacio (IAFE). Buenos Aires, Argentina\label{CONICET}
 \item INAF -- Osservatorio Atrofisico di Catania, via S. Sofia 78, 95123, Catania, Italy\label{catania}
 \item Departamento de F\'isica e Astronomia, Faculdade de Ciencias, Universidade do Porto, Rua Campo Alegre, 4169-007 Porto, Portugal\label{UPorto}
 \item Institut d'Astrophysique de Paris, UMR7095 CNRS, Universite Pierre \& Marie Curie, 98bis boulevard Arago, 75014 Paris, France\label{IAP}
 \item Aix Marseille Univ, CNRS, OHP, Observatoire de Haute Provence, Saint Michel l'Observatoire, France\label{OHP}\\
\end{affiliations}

\begin{bibunit} %[naturemag]

\begin{abstract}
The Earth, Venus, Mars, and some extrasolar terrestrial planets\cite{dressing2015mass} have a mass and radius that is consistent with a mass fraction of about 30\% metallic core and 70\% silicate mantle\cite{stacey2005high}. At the inner frontier of the solar system, Mercury has a completely different composition, with a mass fraction of about 70\% metallic core and 30\% silicate mantle\cite{smith2012gravity}. Several formation or evolution scenarios are proposed to explain this metal-rich composition, such as a giant impact\cite{benz2008origin}, mantle evaporation\cite{cameron1985partial}, or the depletion of silicate at the inner-edge of the proto-planetary disk\cite{wurm2013photophoretic}. These scenarios are still strongly debated. Here we report the discovery of a multiple transiting planetary system (K2-229), in which the inner planet has a radius of 1.165$\pm$0.066~\Rearth\ and a mass of 2.59$\pm$0.43~\Mearth. This Earth-sized planet thus has a core-mass fraction that is compatible with that of Mercury, while it was expected to be similar to that of the Earth based on host-star chemistry\cite{thiabaud2015elemental}. This larger Mercury analogue either formed with a very peculiar composition or it has evolved since, e.g. by losing part of its mantle. 
%This planet, whose day-side temperature can reach up to 2330~K, is an important laboratory to study the mechanisms behind Mercury's composition.
%Together with the forthcoming in-situ mission to Mercury (i.e. BepiColombo), the exploration of Mercury-like exoplanets will provide new constraints on our understanding of the formation of the Solar system's terrestrial planets as well as close-by planets around other stars.
Further characterisation of Mercury-like exoplanets like K2-229~b will help putting the detailed in-situ observations of Mercury (with Messenger and BepiColombo\cite{benkhoff2010bepicolombo}) into the global context of the formation and evolution of solar and extrasolar terrestrial planets.
\end{abstract}

The star EPIC228801451 (TYC 4947-834-1 -- 2MASS J12272958-0643188 -- K2-229) was observed in photometry as part of campaign 10 of the \textit{K2} mission with the \textit{Kepler} space telescope, from 2016-07-06 to 2016-09-20 with a 30-min cadence. Analysis of the extracted and reduced light-curve using the POLAR pipeline\cite*{barros2016new} revealed two sets of periodic planetary transit-like events on periods of about 14~h and 8.3~d. A single transit-like event near the mid-campaign time was also detected (see Fig. 1 and methods). We refer to these planets as K2-229~b, c, and d, respectively. The light curve exhibits a large modulation with a 2\% peak-to-peak amplitude and 18-d period variability (see Fig. 1) that is caused by the presence of active regions (spots and/or faculae). The star is a bright (V=11) late-G / early-K dwarf\cite*{huber2016k2}, hence suitable for precise radial-velocity (RV) observations.

We observed EPIC228801451 with the HARPS spectrograph with the aim of confirming the planetary nature of the transiting candidates and measuring the mass through Doppler spectroscopy. We collected 120 RVs from 2017-01-26 to 2017-05-04 with up to four observations each night. We reduced the spectra using the online pipeline available at the telescope and derived the RV, the full width half maximum (FWHM) and bisector (BIS) of the averaged line profile as well as the spectroscopic indices of chromospheric activity in the core of five spectral lines (see methods and the Supplement Tables 3 and 4). The pipeline automatically rejected one poor-quality spectrum.

To assess the planetary nature of the detected transit signals and rule-out the presence of background stellar objects contaminating the light curve, we performed high-resolution imaging observations with the AstraLux lucky-imaging instrument. No background nor stellar companion is detected within the sensitivity limits of the data and within the photometric mask (see methods and Supplement Fig. 2).

We co-added the HARPS spectra and derived the spectral parameters of the host star\cite{santos2013sweet} (see methods). We find that the host star has an effective temperature of \teff\ = $5120\pm39$~K, a surface gravity of \logg\ = $4.51\pm0.12$~[cgs], an Iron abundance of \met\ = $-0.06\pm0.02$~dex, and a micro-turbulence velocity of v$_{mic}$ = $0.74\pm0.08$~\kms. This identifies the host star as a K0 dwarf. We also derived the chemical abundance of the star (see the Supplement Table 2). 

The 119 RV time series is displayed in Fig. 1 and exhibits a large variability at the level of about 50 \ms\ and a recurrence timescale of 18 days. This large modulation is not caused by planet reflex motions but is due to the presence of active regions on the surface of the star\cite{dumusque2014soap}. We performed a joint analysis of the photometric and spectroscopic data, together with the spectral energy distribution (SED) of the star with the \texttt{PASTIS} software\cite{diaz2014pastis} (see methods). We used three different approaches to correct the RV data for the activity signal of the star\cite{dumusque2017radial}: (I) a nightly RV offset, (II) a Gaussian process (GP) regression, and (III) diagnostics de-correlation with a moving average. For the inner transiting planet (K2-229~b), the three methods give consistent results (see Supplement Fig. 3). Its mass determination is therefore robust in regards to stellar activity. The reflex motion of planet c is only significantly detected with method III and thus requires further investigations to be fully secured. The outer transiting planet (planet d) is not significantly detected in the RV data with any of the methods. We finally adopted the physical parameters of the K2-229 system, that are reported in Table 1, using model II for the correction of the activity-induced RVs.

The inner transiting planet, K2-229~b, has a mass of $2.59\pm0.43$~\Mearth\ and a radius of $1.165\pm0.066$~\Rearth, hence a bulk density of $8.9\pm2.1$~\gcm3. The planets c and d have masses lower than 21.3~\Mearth\ and 25.1~\Mearth\ (95\% credible interval), and a radius of $2.12\pm0.11$~\Rearth\ and $2.65\pm0.24$~\Rearth, respectively.  Note that for the outer transiting planet, K2-229~d, we find two different orbital solutions: with an orbital period close to 31 days or longer than about 50 days. This bi-modal distribution is due to the gap in the first half of the \textit{K2} light curve (see Fig. 1). In the first case, another transit should have occurred during this gap while in the latter case, there is only one transit event during the entire photometric campaign. We excluded the long-period scenario for stability reasons, as this requires large eccentricity to explain the relatively short transit duration ($2.65^{_{+0.15}}_{^{-0.20}}$~h, see Supplement Table 6), making planet d likely to cross the orbit of planet c. 

In Fig. 2, we compare K2-229 b with other known Earth-sized planets\cite{pepe2013earth,berta2015rocky,gillon2017seven,2018arXiv180201377G}, together with theoretical compositions of terrestrial planets (assuming different fractions of metallic core, silicate mantle, and water layer)\cite{brugger2017constraints}. A comparison between K2-229~b and large rocky exoplanets as well as the rocky planets and major moons in the Solar system is also available in the Supplement Fig. 7. Given its high density of $8.9\pm2.1$~\gcm3, we can constrain the water-mass fraction to be less than a few percent at maximum and thus negligible in the planet-mass budget. Assuming K2-229~b is composed only of a metallic core (with a negligible fraction of Si) and a silicate mantle, its mass and radius are consistent with a core-mass fraction (CMF) of $68^{_{+17}}_{^{-25}}$\%\cite{brugger2017constraints}. The derived CMF differs from the one computed ($27^{_{+9}}_{^{-13}}$\% see Fig. 3) assuming the planet was formed with the same Fe/Si$_{c}$ ratio (see methods for a description of Fe/Si$_{c}$) as observed in the host star (Fe/Si$_{c}$=$0.71\pm0.33$). Therefore, the composition of this planet is likely (with an 88\% credible probability) different from the one expected based on the chemical composition of its host star. The reasons for this would be that the planet evolved since its formation, for instance by losing parts of its mantle\cite{cameron1985partial,benz2008origin}, or it was formed with a substantially different Fe/Si$_{c}$ ratio than its host star\cite{wurm2013photophoretic}. With such a potentially large core-mass fraction, K2-229~b stands out as a larger analogue of Mercury in the solar system, which has a CMF of 68\%\cite{stacey2005high,smith2012gravity}. Both planets might therefore share a common formation and evolution history. 

Compared to Mercury, K2-229~b orbits much closer to its star, with an orbital period of about 14~h. Its day-side temperature can reach up to 2330~K, assuming synchronous rotation. These conditions make K2-229 b potentially more sensitive than Mercury to mantle evaporation. At this temperature, the mantle in the day-side of the planet is expected to volatilise into a saturated atmosphere of silicate vapour\cite{leger2011extreme}. Substantial evaporation of this atmosphere would result in planetary mass-loss. Current thermal escape and stellar X-ray irradiation are however not expected to substantially strip out this thin atmosphere (see methods). Other extremely hot rocky worlds have been reported with masses and radii consistent with an Earth-like composition\cite{dressing2015mass}, while a temperate, likely metal-rich, rocky planet has been recently detected: LHS 1140~b\cite{2017Natur.544..333D}. Therefore, stellar irradiation is not expected to contribute to the mass-loss rate (hence mantle evaporation) by more than a few percent of the total planet mass. 

Since K2-229~b is orbiting extremely close to an active K dwarf (at 0.012AU), another hypothesis is that the thin layer of silicate vapour might escape from the planet through magnetic interaction with the host star\cite{lanza2013star, strugarek2016assessing}. Intense stellar wind and flares might also erode the planet's atmosphere\cite{2017arXiv170604617G}. However, these two mechanisms would be less efficient if the planet has its own magnetosphere, protecting the atmosphere from evaporation. More observations and modelling, which are outside the scope of this paper, are required to fully constrain this evolution scenario. The detection of a cometary-like tail\cite{mura2011comet} in the planet vicinity would be an evidence for these mechanisms evaporating the volatilised mantle of rocky planets. Searching for correlation between the CMF of rocky exoplanets and their magnetic environment\cite{moutou2016magnetic} would provide insights on the importance of the magnetic field in this possible mechanism.

Another scenario to explain the composition of Mercury is a Theia-like giant impact\cite{benz2008origin}. If this scenario is responsible for Mercury-like exoplanets, there might be some correlation between the multiplicity and architecture of the systems with the presence of such planets. In that case, these planets would also be prime targets to search for relatively large exo-Moons. However, more giant-impact modelling is needed to understand what would be the initial conditions, in terms of e.g. mass and velocity, for a Theia-like object to remove the mantle of rocky planets much more massive than Mercury such as K2-229~b.

Finally, if photophoresis is the mechanism that forms Mercury-like planets\cite{wurm2013photophoretic}, the comparison of their host star properties compared with those hosting Earth-like planets\cite{dressing2015mass} will provide important constraints on the conditions required for this process. However, given the properties of the Mercury- and Earth-like planets (the orbital separation versus the CMF), there is no clear correlation between the formation conditions and the composition of the system-innermost planet. One possibility would be that the Earth-like planets are formed further out in the disk than those with a Mercury-like composition. Further theoretical work are needed to fully understand the formation and migration of these planets under the photophoresis scenario.

Note that after submitting this paper, the planet K2-106~b, which was initially reported with an Earth-like composition\cite{sinukoff2017k2}, turns out to be much more rich in Iron\cite{guenther2017k2}. No clear comparison between the composition of K2-106~b and the one of its host star has been performed so far. Still, it shows that the existence of such Mercury-like planets is more common than previously thought. This is opening a new approach to understand the particular formation and evolution of these planets.

The planet K2-229~b appears to have abundances that are different from the ones of the the central star. This is the first time such situation is observed in an extrasolar system\cite{dorn2017generalized}. Its composition is similar with that of Mercury. However both planets do not share the same environmental conditions, the former being substantially hotter and closer to its star. Therefore, K2-229~b is an excellent laboratory to test the conditions to form Mercury analogues. The exploration of this new population of exoplanets can greatly complement Solar-system \textit{in-situ} missions like Messenger and BepiColombo\cite{benkhoff2010bepicolombo}, and will help to constrain the formation and evolution mechanisms of Mercury-like planets in various environments. It can then be used to refine the formation models of the Solar-system terrestrial planets. The increased precision in mass and radius measurements brought by upcoming dedicated missions like PLATO\cite{rauer2014plato} will bring an important advancement for this purpose.

%\putbib[EPIC-1451]

\end{bibunit}

\newpage

%\newbibstartnumber{30}
\begin{bibunit} %[naturemag]
\begin{methods}

\subsection{\textit{K2} light-curve reduction and transit detection.}

The search for candidates followed a process incorporating automated searches and human vetting. The POLAR-detrended$^{\citenum{barros2016new}}$ \textit{K2} light curves are first flattened using a moving 3rd-degree polynomial filter, iteratively fit 20 times to a 3.5 day region while ignoring outliers from the previous fit. The polynomial is not allowed to fit across gaps in the data greater than 1 day. After flattening, light curves are run through a modified box-fitting least squares (BLS) search\cite{kovacs2002box}, where outliers across the campaign are ignored from all light curves. We rank light curves on their BLS signal strength, then individually vet every light curve through human eyeballing. Candidates are passed on for follow up if they pass a number of tests: there must be no visible nearby companions, no visible centroid motion during transit and no significant depth difference between odd and even transits. We further rank successful candidates using their transit shape, as measured by the $\theta_2$ statistic\cite{armstrong2017transit}, which uses self-organising maps to distinguish between planetary and false positive transit shapes.

\subsection{Radial velocity observations and reduction.}

The RV observations were carried out with the HARPS spectrograph, mounted on the ESO-3.6-m telescope in La Silla Observatory in Chile. HARPS is a fiber-fed $echelle$ spectrograph with a spectral resolution of R$\sim$110 000, optimised for precise RV observations\cite{phase2003setting}. The spectroscopic data were obtained as part of our ESO -- K2 large programme (ESO programme ID: 198.C-0169). We used an exposure time of 1800s for individual observations, which led to signal-to-noise ratios at the level of 60 per CCD pixel at 550nm . Spectra were reduced with the online pipeline. One spectrum failed the automatic quality checks of the pipeline due to an incorrect colour-flux correction. The RVs were derived following the standard procedure used for more than two decades\cite{baranne1996elodie}, consisting of cross-correlating the observed spectra with a numerical mask corresponding to a G2V star. From the result of the cross-correlation function, we derived the RV, the BIS, the FWHM, and their associated uncertainties\cite{santerne2015pastis,bouchy2001fundamental}. Radial velocities have a median precision of 1.7~\ms. From the spectra, we also measured indices of chromospheric emission in the activity-sensitive \ion{Ca}{2} H\&K (S$_{\rm MW}$), H$_{\alpha}$, \ion{Na}{1} D, and \ion{He}{1} D3 lines\cite{gomes2011long}. All these data are available in Supplement Tables 3 and 4.

From the FWHM values, we estimated the stellar rotational velocity is \vsini = $2.4\pm0.5$~\kms. Using the S$_{\rm MW}$, which evaluates the level of chromospheric emission in the \ion{Ca}{2} H\&K lines, we computed the value of the logR'$_{\rm HK}$\cite{noyes1984rotation} = $-4.58\pm0.04$. The S$_{\rm MW}$ data clearly show a long-term slope with a value of $-0.14\pm0.02$~yr$^{-1}$. This drift is also observed in the FWHM and $I_{H\alpha}$, see the Supplement Fig. 1, and might reveal a rapidly-evolving or large-amplitude magnetic cycle; the star being less active at the end of the campaign than at the start.

\subsection{Spectroscopic analysis and chemical composition of the star.}

The stellar parameters (\teff, \met, \logg, and v$_{mic}$) were derived from the Doppler-corrected co-added HARPS spectra with the same methods as stars in the SweetCat catalog$^{\citenum{santos2013sweet}}$. This method uses standard LTE analysis with the 2014 version of the code MOOG\cite{sneden1974carbon}. A grid of Kurucz ATLAS9 atmospheres\cite{kurucz1993atlas9} was used as input along with the equivalent widths measured automatically with the version 2.0 of the ARES program\cite{sousa2015ares}. The \logg\ value, known to be systematically biased, was corrected using a calibration based on asteroseismic targets\cite{mortier2014correcting}. The derived values are \teff\ = 5120$\pm$39 K, \logg\ = 4.51$\pm$0.12 \cmss, \met\ = -0.06$\pm$0.02 dex, and v$_{mic}$ = 0.74$\pm$0.08 \kms. These values were then used as priors in the Bayesian joint analysis described below. The adopted values, reported in Table 1, are the posteriors of this joint analysis and are thus consistent with the SED and the stellar density as constrained by the transit.

Elemental abundances were determined with the same approach as the analysis of the HARPS GTO sample\cite{adibekyan2012chemical}. Li abundance was derived with spectral synthesis method also by using the code MOOG and ATLAS atmospheres\cite{mena2014li}. Using the abundance ratio [Y/Mg], we estimated the age\cite{maia2016solar} of the star to be 3.0$\pm$0.9 Gyr.

%In order to derive chemical abundances we first co-added all the Doppler-corrected individual spectra into a single 1D spectra with IRAF. The chemical abundances for all elements, except Lithium, were derived under a  This approach is the same as for the analysis of the HARPS GTO sample where further details and the complete list of lines used can be found. Li abundances were derived with spectral synthesis also using the code MOOG and ATLAS atmospheres\cite{mena2014li}.

\subsection{Lucky Imaging observations.}

We obtained a high-spatial resolution image of the target with the AstraLux camera\cite{hormuth08} installed at the 2.2m telescope of Calar Alto Observatory (Almer\'ia, Spain). By using the lucky-imaging technique to avoid atmospheric distortions, we obtained 90,000 images in the SDSS i$^{\prime}$ band of 30 ms exposure time, well below the coherence time. We used the observatory pipeline to perform the basic reduction of the images and the subsequent selection of the 10\% best-quality frames, with the highest Strehl ratios (total integration time of 270\,s). These images are then aligned and combined to obtain the final high-spatial resolution image, shown in Supplement Fig. 2 together with the 80x80'' image of the Digital Sky Survey (DSS) and the \textit{K2} photometric mask. No other companion closer than 12'' and within the sensitivity limits is detected. We estimated the sensitivity curve\cite{lillo2014high} of this high-spatial resolution image, shown in Supplement Fig. 2.

Due to the de-centering of the \textit{K2} mask compared to the AstraLux image, we also checked for companions in the DSS image and found only a faint target ($\Delta R=8.1$~mag) outside of the \textit{K2} best aperture. The contamination from this source is thus negligible. Consequently, we can conclude that the extracted \textit{K2} lightcurve is not contaminated within our sensitivity limits. 
%We can quantify the probability of having missed a stellar companion that could affect substantially the transit signal of the detected planets. The maximum magnitude contrast that an eclipsing binary could have to mimic the transit signal is given by $-2.5\log\delta$, where $\delta$ corresponds to the transit depth. For the three planets this value corresponds to $\Delta m^{\rm max}_{b} = -9.4$, $\Delta m^{\rm max}_{c} = -8.1$, $\Delta m^{\rm max}_{d} = -7.6$. We used these values as the maximum contrast for contamination (represented by horizontal dashed lines in Fig.~XX). Given this limit, we can estimate the probability that our high-spatial resolution image missed a star brighter than this limit and closer than 6''. In \cite{lillo-box14}, we defined 
We computed the background source confidence (BSC)\cite{lillo2014high}, which illustrates the confidence to which we can affirm that our planet signal is not affected by relevant chance-aligned sources up to a certain separation. In this case, we obtained BSC values of 98.32\%, 99.44 \%, and 99.67 \%, for planets b, c, and d, respectively. Therefore, we can conclude that the chances of having missed a relevant contaminant in this high-spatial resolution image are very small for any of the detected planets.

\subsection{Bayesian analysis of photometry, RV, and SED.}

We jointly analysed the \textit{K2} photometry, HARPS RVs and the SED in a Bayesian framework. We used the K2SFF reduction\cite{vanderburg2014technique} of the \textit{K2} light curve with the optimal aperture mask that we corrected for spot and faculae modulation as well as residuals of instrumental systematics, using a Gaussian process regression with a Mat\`ern 3/2 kernel\cite{rasmussen2006gaussian}, trained on out-of-transit photometry. For the SED, we used the optical magnitudes from the APASS survey\cite{henden2014apass}, and near-infrared magnitudes from the 2-MASS and WISE surveys\cite{cutri2014vizier}. These magnitudes are reported in Supplement Table 1. The light curve was modelled with the \texttt{jktebop} software\cite{southworth2008homogeneous} with an oversampling factor of 30 to account for the long exposure time of \textit{Kepler}\cite{kipping2010binning}. The SED was modelled with the BT-STTL stellar atmosphere models\cite{allard2012models}. For the RV, we used three different models, described below, to account for the Keplerian orbit of the three planets and the stellar variability. We did not oversample the RV model to account for the relatively long exposure time (3.5\% of period of the inner planet) as this effect is expected to under-estimate the planet mass by less than 0.2\% which is negligible.

To account for the asterodensity profiling\cite{kipping2014characterizing}, we modelled the central star with the Dartmouth stellar evolution tracks\cite{dotter2008dartmouth}. Note that we also used the PARSEC evolution tracks\cite{bressan2012parsec} with no significant change in the result. Linear and quadratic limb-darkening coefficients were taken from a commonly adopted theoretical table\cite{claret2011gravity} and are changed in the analysis as a function of the stellar parameters.

The analysis was performed using the \texttt{PASTIS} software$^{\citenum{diaz2014pastis}}$ which runs a Markov Chain Monte Carlo (MCMC) algorithm. The exhaustive list of parameters and their respective priors is available in the Supplement Tables 5--7 and displayed in Supplement Fig. 4--6. For each of the activity-correction methods, we run 20 MCMC chains of 3$\times$10$^{5}$ iterations, initialised at random points drawn from the joint prior distribution. The chains were tested for convergence using a Kolmogorov--Smirnov test\cite{brooks2003nonparametric}. The burn-in of each chain was removed before merging them to derive the posterior distribution from which we computed the median and 68.3\% credible interval for each parameters. When necessary, we kept only the samples of the posterior distributions for which the stellar age, as provided by the evolution tracks, is compatible with the age of the Universe. Similar analyses have previously been performed to characterise several transiting companions, from brown dwarfs to low-mass planets\cite{santerne2014sophie, bayliss2016epic, osborn2017k2}.

\subsection{Activity-correction method I: a nightly RV offset.}

A straightforward method to correct for stellar activity when characterising ultra-short period planets is to account for a nightly RV offset\cite{hatzes2011mass}. This requires that several observations are taken each night which is the case here. This method assumes that the activity of the star does not evolve substantially within the night, but filters out any signal with a period longer than about one day. Therefore, this model can only be used to characterise the inner transiting planet, K2-229~b.

The function $\mathbf{f}$ to describe the RV data is the following:
\begin{eqnarray}
\mathbf{f} &=& \gamma + \mathbf{k_{j=1}} + \bm{\Delta\gamma_{l}}\ ,\\
\mathbf{k_{j}} &=& K_{j}\cdot\left[ \cos\left(\bm{\nu_{j}} + \omega_{j}\right) + e_{j}\cos(\omega_{j})\right] \,
\end{eqnarray}
where $\gamma$ is the systemic RV, $\mathbf{k_{j}}$ is the Keplerian orbit model of the $j^{th}$ planet, and $\bm{\Delta\gamma_{l}}$ is the offset of the $l^{th}$ night. The Keplerian orbit of the planet $j$ is described by $K_{j}$ the RV semi-amplitude, $\bm{\nu_{j}}$ the true anomaly, $e_{j}$ the eccentricity, and $\omega_{j}$ the argument of periastron.

The target K2-229 was observed over 51 different nights with at least two and up to four observations. For the sake of simplicity, we here fixed the ephemeris of the inner planet to their best values (see supplement table 5). We find no significant eccentricity, with $e_{1} = 0.16\pm0.11$, as expected given the ultra-short period of the planet. Assuming a circular orbit, we find $K_{1} = 2.82\pm0.40$ \ms, which translates to a planetary mass of M$_{b}$ = $3.28\pm0.47$~\Mearth. Note that the RV activity signal varies up to 1.1~\ms\ over the $\sim$5~h observations each night. This mass determination might thus be slightly biased depending on how the activity signal phases with the planet orbit at the time of the observations.

\subsection{Activity-correction method II: Gaussian process regression.}

Recently, GPs have been shown to be a robust statistical method to model stellar activity in RV data$^{\citenum{dumusque2017radial},}$\cite{haywood2014planets}. For that, we computed the likelihood $\mathcal{L}$ accounting for the covariance matrix\cite{rasmussen2006gaussian}.
\begin{eqnarray}
\ln\mathcal{L} &=& -\frac{1}{2}\mathbf{r}^{\top}\mathbf{K}^{-1}\mathbf{r} - \frac{1}{2}\ln\left|\mathbf{K}\right| - \frac{n}{2}\ln(2\pi)\ ,\\
\mathbf{r} &=& \mathbf{y} - \left[\gamma + \sum_{j=1}^{3}\mathbf{k_{j}}\right]\ ,\\
\mathbf{K} &=& A^{2}\exp\left[-\frac{1}{2}\left(\frac{\mathbf{\Delta t}}{\lambda_{1}}\right)^{2} - \frac{2}{\left(\lambda_{2}\right)^{2}}\sin^{2}\left(\frac{\pi\mathbf{\Delta t}}{P_{rot}}\right)\right] + \mathbf{I}\sqrt{\bm{\sigma}^{2}+\sigma_j^{2}}\ ,\label{QPkernel}
\end{eqnarray}
where $n$ is the number of observations, $\mathbf{r}$ is the residual between the three Keplerian-orbit model and the data $\mathbf{y}$, and $\mathbf{K}$ is the covariance matrix. We used here a \textit{quasi-periodic} kernel\cite{rasmussen2006gaussian} (eq. \ref{QPkernel}) with the hyper-parameters $A$ the amplitude, $P_{rot}$ the rotation period, $\lambda_{1}$ the coherent timescale of active region, and $\lambda_{2}$ the relative weight between the periodic and decay terms. The matrix $\mathbf{\Delta t}$ is the differential time such that $\mathbf{\Delta t_{ij}} = t_{i} - t_{j}$. Finally, $\mathbf{I}$ is the identity matrix, $\bm{\sigma}$ is the Gaussian uncertainty vector associated to the data and $\sigma_j$ is an extra source of uncorrelated (jitter) noise.

We first trained the GP hyper-parameters on the \textit{K2} light curve. However, because of the 15d gap in the first part of the campaign, the rotation period was only constrained to $P_{rot}$ = $18.1\pm1.1$~d. We also trained the GP hyper-parameters on the spectroscopic diagnoses. We find $P_{rot}$ of $18.1\pm0.3$~d, $19.2\pm0.4$~d, $18.1\pm0.2$~d, $18.4\pm0.3$~d, $17.1\pm0.4$~d, and $17.9\pm1.0$~d by training on the S$_{\rm MW}$, FWHM, BIS, $I_{H\alpha}$, $I_{NaD}$, and $I_{HeID3}$, respectively. All these estimates of $P_{rot}$ are fully consistent and we finally adopted the one based on the S$_{\rm MW}$, hence $P_{rot}$ = $18.1\pm0.3$~d.

We analysed the HARPS RVs of K2-229 with a three Keplerian-orbit model and a GP as described above. We used uninformative priors for the hyper-parameters except for $P_{rot}$ for which we use a normal distribution of $18.1\pm0.3$~d.

We first allowed the eccentricity of the three planets to vary. We find non-significant eccentricities with $e_{1}$ = 0.08$^{_{+0.08}}_{^{-0.06}}$, $e_{2}$ = 0.30$^{_{+0.36}}_{^{-0.13}}$, and $e_{3}$ = 0.32$\pm$0.31. We thus assumed circular orbits for the two inner planets which have orbital periods of less than 10 days and are likely circularised. We find RV amplitudes of $K_{1} = 2.23 \pm 0.35$~\ms, $K_{2} = 3.2 \pm 2.4$~\ms, and $K_{3} = 2.4^{_{+2.8}}_{^{-1.6}}$~\ms, for planets b, c, and d (respectively). This corresponds to a mass for planet b of M$_{b} = 2.59 \pm 0.43$~\Mearth. Planet c and d only have upper-limits on their masses of 21.3~\Mearth\ and 25.1~\Mearth\ with a 95\% credible probability, respectively. The RV residuals from the best model have a scatter of 1.6~\ms, which is fully compatible with the median photon noise on this star (1.7~\ms).We therefore adopted this solution for the system.

\subsection{Activity-correction method III: de-correlation of the spectroscopic diagnoses and moving average.}

The last approach to correct for stellar activity in RV data consists in using the spectroscopic diagnoses (e.g. FWHM, BIS, S$_{\rm MW}$, $I_{H\alpha}$)\cite{anglada2015comment}. This technique then assumes a linear correlation between the RVs and each of the diagnoses. The coefficients of the linear correlation are fitted simultaneously with the Keplerian orbit models. The model $f$ for each observing time $t_{i}$ described below also accounts for a moving average with an exponential decay in time. This method was shown to be efficient to correct for activity-induced RV variation in the recent RV challenge$^{\citenum{dumusque2017radial}}$.

\begin{eqnarray}
f(t_{i}) &=& \gamma + at_{i} + bt_{i}^{2} + \sum_{j=1}^{3}k_{j}(t_{i})\nonumber\\
&& + c_{0}\left(BIS(t_{i}) - BIS_{0}\right) + c_{1}\left(FWHM(t_{i})-FWHM_{0}\right) \nonumber\\
&& + c_{2}\left(S_{\rm MW}(t_{i}) - S_{{\rm MW},0}\right) + c_{3}\left(I_{H\alpha}(t_{i}) - I_{H\alpha,0}\right) \nonumber\\
&& +\phi\left[f(t_{i-1}) - \sum_{j=1}^{3}k_{j}(t_{i-1})\right]\exp\left(\frac{t_{i-1}-t_{i}}{\tau}\right)\ ,
%k_{j}(t_{i}) &=& K_{j}\cdot\left[ \cos\left(\nu_{j}(t_{i}) + \omega_{j}\right) + e_{j}\cos(\omega_{j})\right]
\end{eqnarray}
where $a$, $b$ are coefficients for a possible linear and quadratic (respectively) long-term drift, $c_{n}$ are the coefficients for the linear correlation with the spectroscopic diagnoses. The parameters $BIS_{0}$, $FWHM_{0}$, $S_{{\rm MW},0}$, and $I_{H\alpha,0}$ are the zero point of the corresponding diagnoses. The moving average has two parameters, $\phi$ its amplitude  and $\tau$ its time decay.

The spectroscopic diagnoses $I_{NaD}$ and $I_{HeID3}$ were not used in this model because the activity signal is only barely detected in these data. Note we tested the influence of each diagnoses and the best solution is found when the moving average is used together with the BIS, FWHM, and S$_{\rm MW}$ indices. Including $I_{H\alpha}$ or not in the model does not significantly change the solution, as this diagnosis is redundant with S$_{\rm MW}$ at our level of precision. This implies that $c_{2}$ and $c_{3}$ are fully degenerated. Nevertheless, we included $I_{H\alpha}$ in the analysis as this has no significant effect on the result.

As for the previous methods, we first let the eccentricity of the three planets vary in the analysis and find that $e_{1}$ = $0.20\pm0.11$, $e_{2}$ = $0.24\pm0.12$, and $e_{3}$ = $0.49\pm0.39$. Since the orbit of the two inner planets, which are expected to be circularised, have no significant eccentricity, we then assumed them as circular. We find RV semi-amplitudes of $K_{1}$ = $2.11\pm0.73$~\ms, $K_{2}$ = $3.33\pm0.85$~\ms, and $K_{3}$ = $1.1^{_{+1.4}}_{^{-0.7}}$~\ms, for planets b, c, and d, respectively. This translates into a mass for planets b and c of M$_{b}$ = $2.45\pm0.84$~\Mearth\ and M$_{c}$ = $9.3\pm2.4$~\Mearth, respectively. For the planet d, we only measured an upper-limit on the mass at 11.5~\Mearth\, with a 95\% credible probability. These values are fully compatible with the ones derived with the other methods. This method gives an uncertainty for planet b twice as large as the two other methods but allows us to tentatively measure the mass of planet c. The RV residuals of the best model have a scatter at the level of 5.5~\ms, which is significantly higher than the median photon noise (1.7~\ms). Thus, this model does not fully correct the activity-induced RVs for this star.

\subsection{Likelihood of planetary signals.}

We have significantly detected the reflex motion of the host star caused by the orbit of planet b, with three different methods to correct for stellar activity. The spectroscopic diagnoses which might reveal false-positive scenarios\cite{santerne2015pastis} show no significant variation at the orbital period of planet b (with false-alarm probabilities larger than 50\%) in the GP-corrected data (see Supplement Fig. 1). Moreover, if the RV variations caused by planet b were correlated with the BIS or FWHM, method III would have likely absorbed its Keplerian signal. Moreover, the AstraLux high-resolution imaging revealed no contaminating stars within the sensitivity limits. We therefore conclude that planet b is a \textit{bona-fide} transiting planet. 

The Keplerian signal of planet c is detected only with method III. Once again, if this RV signal was correlated with spectroscopic diagnoses, it would have been unlikely to be detected with this method. Using the planet-multiplicity likelihood boost\cite{lissauer2014validation}, we consider planet c is also a \textit{bona-fide} transiting planet. 

With only a unique and partial transit and no detection in the RV data, planet d is not fully secured. More photometric observations are needed to confirm the presence of this third planet in the system. If this mono-transit event is real, given that two planets are already transiting in this system and no background star is detected in the photometric mask within the AstraLux sensitivity, planet d would likely be a \textit{bona-fide} transiting planet. 

\subsection{Planetary composition.}

To probe the composition of K2-229 b, we used a model of planetary interior models based on the physical properties of the Earth and terrestrial planets$^{\citenum{brugger2017constraints}}$. All planets considered in this model are fully differentiated into three main layers: a metallic core, a silicate mantle, and a water envelope. In the case of K2-229 b, given its high density, we can constrain from our simulations that the water-mass fraction (WMF) is less than about 10\% if in liquid phase, with almost no mantle. Assuming water in super-critical or gaseous phase decreases the WMF to a few percent only at maximum. Thus, we only consider solid terrestrial structures in this work. In our model, the core is composed of iron Fe and the iron alloy FeS in proportions 87\% to 13\%, as in the case of the Earth\cite{sotin2007mass}. We note that other volatiles (e.g. O, Si, C) are commonly used to model the Earth's core\cite{badro2014seismologically} with abundances up to a few percent. For a sake of simplicity, here, we only consider sulfur in our model ; it is assumed as a proxy of all volatiles.

The silicate mantle might be divided into two sublayers: a lower mantle composed of bridgmanite and ferro-periclase, and an upper mantle composed of olivine and enstatite, due to a a phase transition induced by the pressure gradient inside the planet. The size and thus mass of the core and mantle fix the distribution of materials inside a planet, i.e. its composition. From this, a planet's composition is entirely described by a single parameter, namely its core-mass fraction (CMF). In our simulations, we assumed the surface conditions of K2-229 b to be 1~bar pressure, with a surface temperature equal to the derived equilibrium temperature. At this temperature, the silicate mantle would be liquid on the surface of the planet, however we do not consider here this phase transition, as the density of liquid silicates differs from that of the solid phase by only a few percent$^{\citenum{leger2011extreme},}$\cite{lebrun2013thermal}. The model computes a planetary radius from a given planetary mass and an assumed composition (or CMF). We thus explored the full 0--100\% range for the CMF, and a limited range for the planetary mass (see Figure~3), in order to constrain the composition of K2-229 b from its measured fundamental parameters.

For terrestrial planets, the CMF is directly linked to the bulk Fe/Si and Mg/Si ratios$^{\citenum{brugger2017constraints}}$. As these ratios cannot be measured, we assumed them to be consistent to the stellar values$^{\citenum{thiabaud2015elemental},\citenum{dorn2017generalized}}$. Using the stellar abundances from Supplement Table~2, that are relative to that of the Sun\cite{lodders2010solar}, we find Fe/Si = $0.70\pm0.33$ and Mg/Si = $0.89\pm0.37$ for K2-229. For terrestrial planets, these values can be corrected to include elements that are not considered in the model (Al, Ca, and Ni) but represent a non-negligible part of the planets' mass\cite{allegre1995chemical, sotin2007mass}. Taking into account these elements, we computed corrected ratios Fe/Si$_{c}$ = $0.71\pm0.33$ and Mg/Si$_{c}$ = $0.98\pm0.40$. The CMF of a planet is essentially governed by the Fe/Si ratio, as in our model, Fe is mostly found in the core, whereas Mg and Si are only found in the mantle. As shown in Figure~3, Fe/Si$_{c}$ = $0.71\pm0.33$ corresponds to CMF = $27^{_{+9}}_{^{-13}}$\%. As they are measured relative to the Sun, the derived Fe/Si and Mg/Si absolute ratios of the star can significantly vary depending on the various solar abundances available in the literature\cite{grevesse2007solar,asplund2009chemical}. Using different values for the solar abundances, we find that the difference between the expected CMF, derived from the stellar Fe/Si$_{c}$, and the one derived from the fundamental parameters of K2-229 b, remains: Fe/Si$_{c}$ = $0.82\pm0.25$ (hence CMF = $26^{_{+6}}_{^{-8}}$\%)\cite{asplund2009chemical} or Fe/Si$_{c}$ = $0.74\pm0.26$ (hence CMF = $28^{_{+7}}_{^{-9}}$\%)\cite{grevesse2007solar}. Therefore, the expected value of the CMF is not sensitive to the values used for the solar abundances.

The validity of these results assumes no inclusion of Si in the planet's core. If Si is actually present in the core in substantial amount, it would change the planetary Fe/Si towards a more stellar value, hence impacting our conclusion. However, in the case of the Earth's core, the fraction of Si is constrained at the level of a few percent at maximum\cite{badro2014seismologically}.

\subsection{Mantle evaporation.}

%The Hertz-Knudsen formula gives the evaporation mass flux of particles, $F_{HK}$, from the surface of a liquid or solid species.
%\begin{equation}
%F_{HK} = \alpha \sqrt{\frac{M}{2\pi R T}}(P_S-P)\ ,
%\end{equation}
%where $P_S$ is the vapour saturation pressure, $P$ is the pressure, $T$ is the temperature, $M$ is the molar mass, $R$ is the universal gas constant and $\alpha$ is a coefficient depending on the species. For enstatite, which is assumed to be the main component of the mantle at the surface, $P_S$ is related to $T$ as follows$^{\citenum{cameron1985partial}}$.
%\begin{equation}
%\log P_S = 13.176 - \frac{24605.0}{T} .
%\end{equation} 
%For enstatite, the coefficient $\alpha$ was estimated$^{\citenum{cameron1985partial}}$ as $\alpha\approx 0.1$. From this, we get the total evaporation rate under zero pressure at $T=2330$~K and integrated over the day-side of the planet of $F_{HK} = 0.2$~\Mearth.Myr$^{-1}$.
	
The volatilisation of the atmosphere might be achieved by stellar irradiation. Considering that the escape is energy-limited, the mass-loss rate $\dot{m}$ is obtained by equalising the gravitational potential with the incident energy flux:
\begin{equation}
\dot{m} = \epsilon \frac{L_{X} R_p^3}{GM_p (2a)^2}\ ,
\end{equation}
where $L_X$ is the X-ray luminosity, $R_p$ and $M_{p}$ are the planetary radius and mass (respectively), $a$ is the semi-major axis, $G$ is the gravitational constant, and $\epsilon$ is an efficiency factor estimated\cite{owen2012planetary} as $\epsilon\simeq 0.12$. We estimated the X-ray flux\cite{pallavicini1981relations} as $L_X \simeq 1.4\cdot (\vsini)^{1.9} = 7.6\cdot 10^{27}$~erg.s$^{-1}$. The mass-loss rate through irradiation is then $\dot{m} = 1.3\cdot 10^{-5}$~\Mearth.Myr$^{-1}$. Note that this is slightly underestimated as the irradiation occurs in the exosphere and not at the planetary surface. This mass-loss rate was also substantially higher in the early stage of the system, when the star was younger and thus more active. Depending on the system's age, this mechanism might have eroded the atmosphere at the level of only a few percent of the total planet mass.
	
The volatilisation of the atmosphere might also be achieved by Jeans (or thermal) escape. In that case, the particle-loss rate is given by the Jeans formula: 
\begin{equation}
F_J = \frac{N_{ex}v_0}{2 \sqrt{\pi}}(1+\lambda_{esc})\exp^{-\lambda_{esc}}\ ,
\end{equation}
where $N_{ex}$ is the particle density at the exobase, $v_0=\sqrt{\frac{2kT}{m}}$ is the mean thermal velocity of a particle of mass $m$ and $\lambda_{esc}$ is the adimensional escape parameter, defined as:
\begin{equation}
\lambda_{esc} = \frac{G M_p m}{k_BT(R_p+H)}\ ,
\end{equation}
with $H$ the exobase height and $k_B$ the Boltzmann constant.

In order to compute $F_J$, we assumed that the gaseous enstatite is completely photo-dissociated in an ideal gas of mean molecular weight $M_{mol} \simeq 11$~g.mol$^{-1}$. Under these conditions, we get a mass-loss rate of $\dot{m} = 5.2\cdot 10^{-33}$~\Mearth.Myr$^{-1}$. Even at the extreme yet possible temperature of 10$^4$~K at the exobase, the mass-loss rate only reaches $2.8\cdot 10^{-14}$~\Mearth.Myr$^{-1}$, which means that this process is not sufficient to explain the high density of K2-229~b.

\end{methods}

\renewcommand{\refname}{Methods references}
%\putbib[EPIC-1451]

\end{bibunit}

%\bibliographystyle{naturemag}
%\bibliography{EPIC-1451}
%\bibliographymain{EPIC-1451}

%% Here is the endmatter stuff: Supplementary Info, etc.
%% Use \item's to separate, default label is "Acknowledgements"

\begin{addendum}
 \item We are grateful to the pool of HARPS observers who conducted part of the visitor-mode observations at La Silla Observatory: Romina Ibanez Bustos, Nicola Astudillo, Aur\'elien Wyttenbach, Esther Linder, Xavier Bonfils, Elodie H\'ebrard, and Alejandro Suarez. A.S. thanks Emannuel Hugot for fruitful comments on the manuscript. We warmly thank the four anonymous reviewers for the time they spent reading and commenting this manuscript that greatly helped improving its quality.

Based on observations collected at the European Organisation for Astronomical Research in the Southern Hemisphere 
under ESO programme 198.C-0168.

This publication makes use of The Data \& Analysis Center for Exoplanets (DACE), which is a facility based at the University of Geneva (CH) dedicated to extrasolar planets data visualisation, exchange and analysis. DACE is a platform of the Swiss National Centre of Competence in Research (NCCR) PlanetS, federating the Swiss expertise in Exoplanet research. The DACE platform is available at https://dace.unige.ch. This research has made use of the NASA Exoplanet Archive, which is operated by the California Institute of Technology, under contract with the National Aeronautics and Space Administration under the Exoplanet Exploration Program. This research has made use of the VizieR catalogue access tool, CDS, Strasbourg, France. The original description of the VizieR service was published in A\&AS 143, 23. This publication makes use of data products from the Two Micron All Sky Survey, which is a joint project of the University of Massachusetts and the Infrared Processing and Analysis Center/California Institute of Technology, funded by the National Aeronautics and Space Administration and the National Science Foundation. This publication makes use of data products from the Wide-field Infrared Survey Explorer, which is a joint project of the University of California, Los Angeles, and the Jet Propulsion Laboratory/California Institute of Technology, funded by the National Aeronautics and Space Administration.

The Porto group acknowledges the support from Funda\c{c}\~ao para a Ci\^encia e a Tecnologia (FCT) through national funds and from FEDER through COMPETE2020 by the following grants: UID/FIS/04434/2013 \& POCI--01--0145-FEDER--007672, PTDC/FIS-AST/1526/2014 \& POCI--01--0145-FEDER--016886, and PTDC/FIS-AST/7073/2014 \& POCI-01-0145-FEDER-016880. 
FCT is further acknowledged through the Investigador FCT contracts IF/01312/2014/CP1215/CT0004 (SCCB), IF/00849/2015/CP1273/CT0003 (EDM), IF/00650/2015/CP1273/CT0001 (VA), IF/01037/2013/CP1191/CT0001 (PF), IF/00169/2012/CP0150/CT0002 (NCS), and IF/00028/2014/CP1215/CT0002 (SGS) and for the fellowships SFRH/BD/93848/2013 (JPF), PD/BD/128119/2016 (SH) and PD/BD/52700/2014 (JJN), that are funded by FCT (Portugal) and POPH/FSE (EC).
%J.P.F., S.H., and J.J.N. acknowledge support by 
%PF further acknowledges support from FCT in the form of an exploratory project of reference IF/01037/2013CP1191/CT0001.
JLB  acknowledges support from the Marie Curie Actions of the European Commission (FP7-COFUND).  DBar has been supported by the Spanish grant ESP2015-65712-C5-1-R.
DJA is funded under STFC consolidated grant reference ST/P000495/1. DJAB acknowledges support from the University of Warwick and the UKSA. EF is funded by the Qatar National Research Foundation (programme QNRF-NPRP-X-019-1).
XD is grateful to the Society in Science -- The Branco Weiss Fellowship for its financial support. RL thanks CNES for financial support through its post-doctoral programme.
The project leading to this publication has received funding from Excellence Initiative of Aix-Marseille University -- A*MIDEX, a French ``Investissements d'Avenir'' program. The French group acknowledges financial support from the French Programme National de Plan\'etologie (PNP, INSU). This work has been carried out in the frame of the National Centre for Competence in Research (NCCR) "PlanetS" supported by the Swiss National Science Foundation (SNSF)

 \item[Competing Interests] The authors declare that they have no
competing financial interests.
 \item[Authors Contributions] The Warwick group (DJA, DJAB, AD, FF, EF, JJ, GK, JK, WKFL, TL, JMC, HPO, DP) has detected the candidates. SCCB, OD, and MD reduced the POLAR -- \textit{K2} light curves used in the candidate search. The Geneva group (DBay, FB, RFD, XD, HG, CL, FP, SU) have organised the HARPS runs, reduced the HARPS data, and developed the DACE tool. ASB, SGS, DJA, and JPF performed part of the visitor-mode runs on HARPS. MH derived the spectroscopic indices. VA, ADM, NCS, SGS have derived the stellar parameters and chemical composition of the star. JLB and DBar have performed the AstraLux observations and derived the background source confidence. HG and AS performed the Bayesian analysis of the data with the \texttt{PASTIS} code that was initially developed by RFD, J-MA, and AS. BB, MD, OM, and AA developed the planet composition model and discussed the mantle evaporation. VA, J-MA, DBar, SCCB, DBay, IB, ASB, FB, DJAB, MD, EDM, OD, JPF, PF, HG, GH, SH, RL, CL, JJN, HPO, FP, DP, NCS, SGS, SU, and AV are CoIs of the ESO -- \textit{K2} large programme that is coordinated by AS who led the ESO proposal. AS led this study and wrote most of the manuscript. All co-authors contributed to the discussion of the paper.
 \item[Data Availability Statement] The data that support the plots within this paper and other findings of this study are available from the corresponding author upon reasonable request.
 \item[Correspondence] Correspondence and requests for materials
should be addressed to A.~Santerne~(email: alexandre.santerne@lam.fr).
\end{addendum}

\newpage

\begin{figure}
\begin{center}
\includegraphics[width=\textwidth]{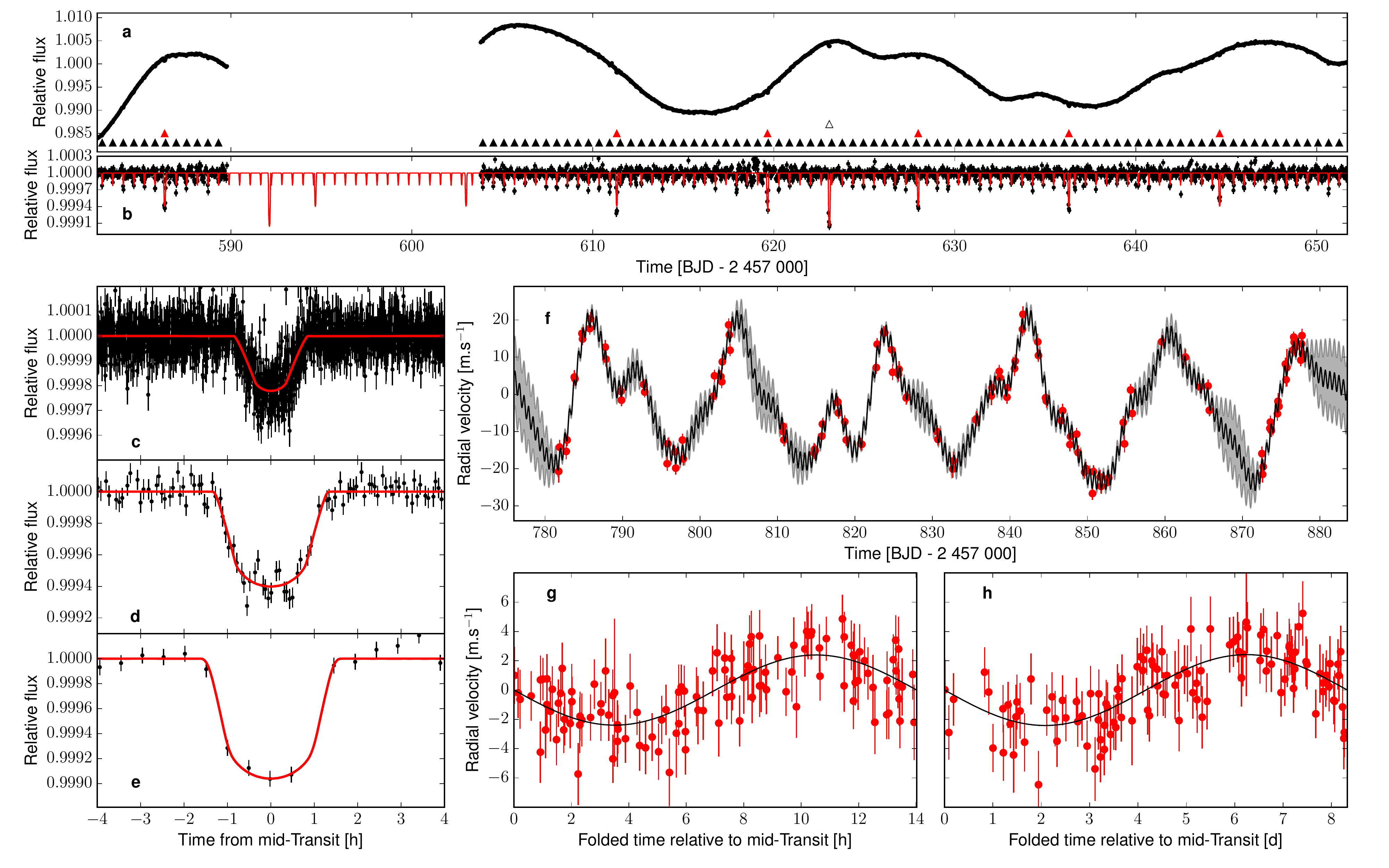}
\label{fig1}
\caption{\textbf{Photometric and RV data of the K2-229 system. (a)} Extracted and normalised \textit{K2} light curve. The times of transit of the planets b, c, and, d are indicated by the black, red, and white triangles (respectively). \textbf{(b)} Gaussian-process flattened light curve (black) with the best 3-planet photometric model (red line). \textbf{(c, d, e)} Phase-folded transit light curves of the planets b, c, and d (respectively). The best transit model is displayed in red. \textbf{(f)} Radial velocity time series obtained with HARPS (in red) together with the best 3-Keplerian orbit and a Gaussian process regression (in black). The grey region corresponds to the 68.3\% credible interval from the Gaussian process. \textbf{(g, h)} Time-folded RV data (in red) to the transit epoch and period of the planets b and c (respectively). The data are corrected from the activity signals and the other planetary contribution. The best Keplerian orbit models are shown in black.}
\end{center}
\end{figure}

\begin{figure}
\begin{center}
\includegraphics[width=\textwidth]{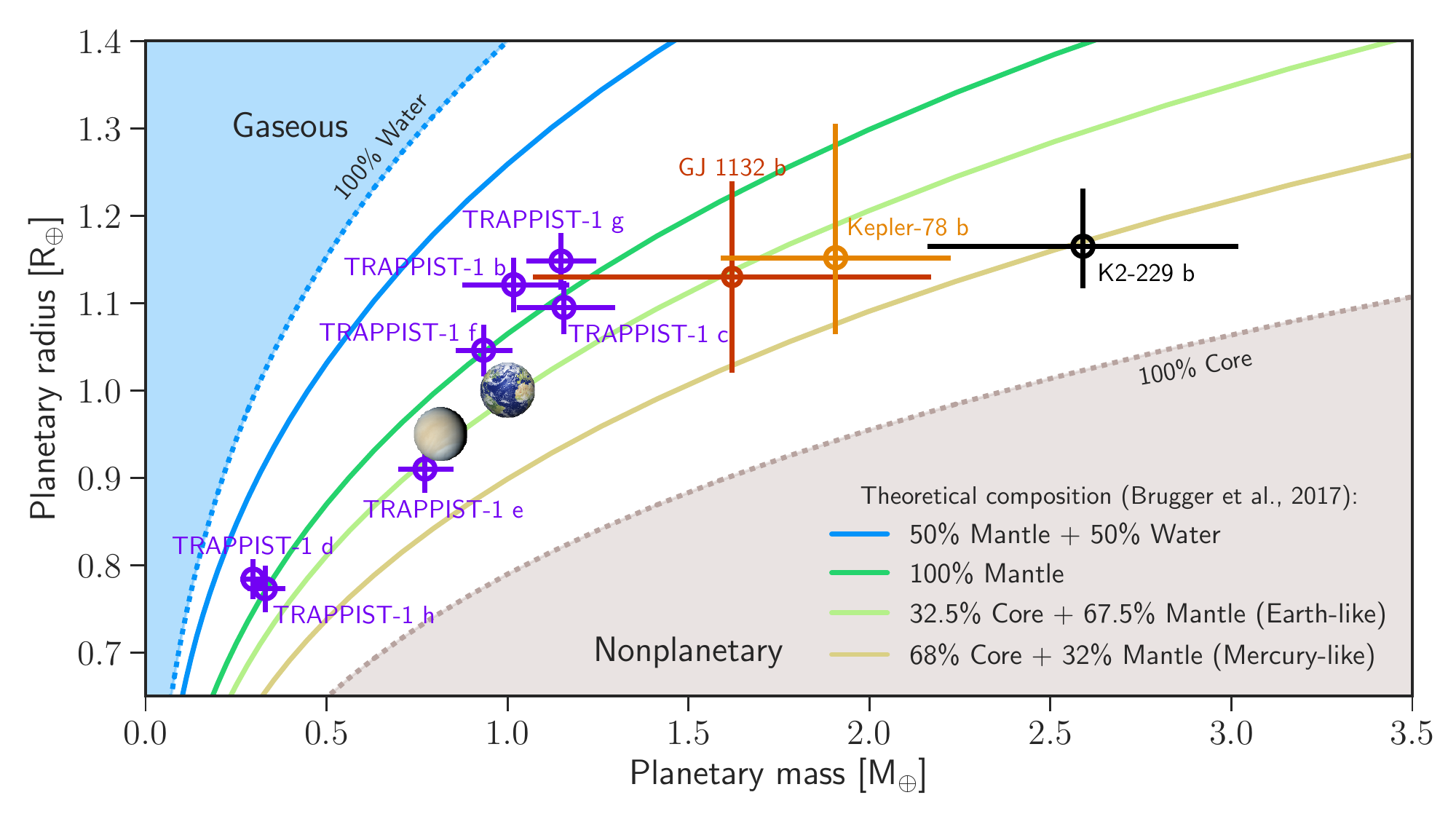}
\label{fig2}
\caption{\textbf{Mass -- Radius diagram of known Earth-sized planets.} Only planets that have a mass measured with a precision better than 50\% are shown here (source: NASA exoplanet archive). The different lines represent possible theoretical compositions for terrestrial worlds$^{\citenum{brugger2017constraints}}$. Objects denser than 100\% core are considered as non-planetary objects and those less dense than 100\% water are considered to be gaseous.}
\end{center}
\end{figure}

\begin{figure}
\begin{center}
\includegraphics[width=\textwidth]{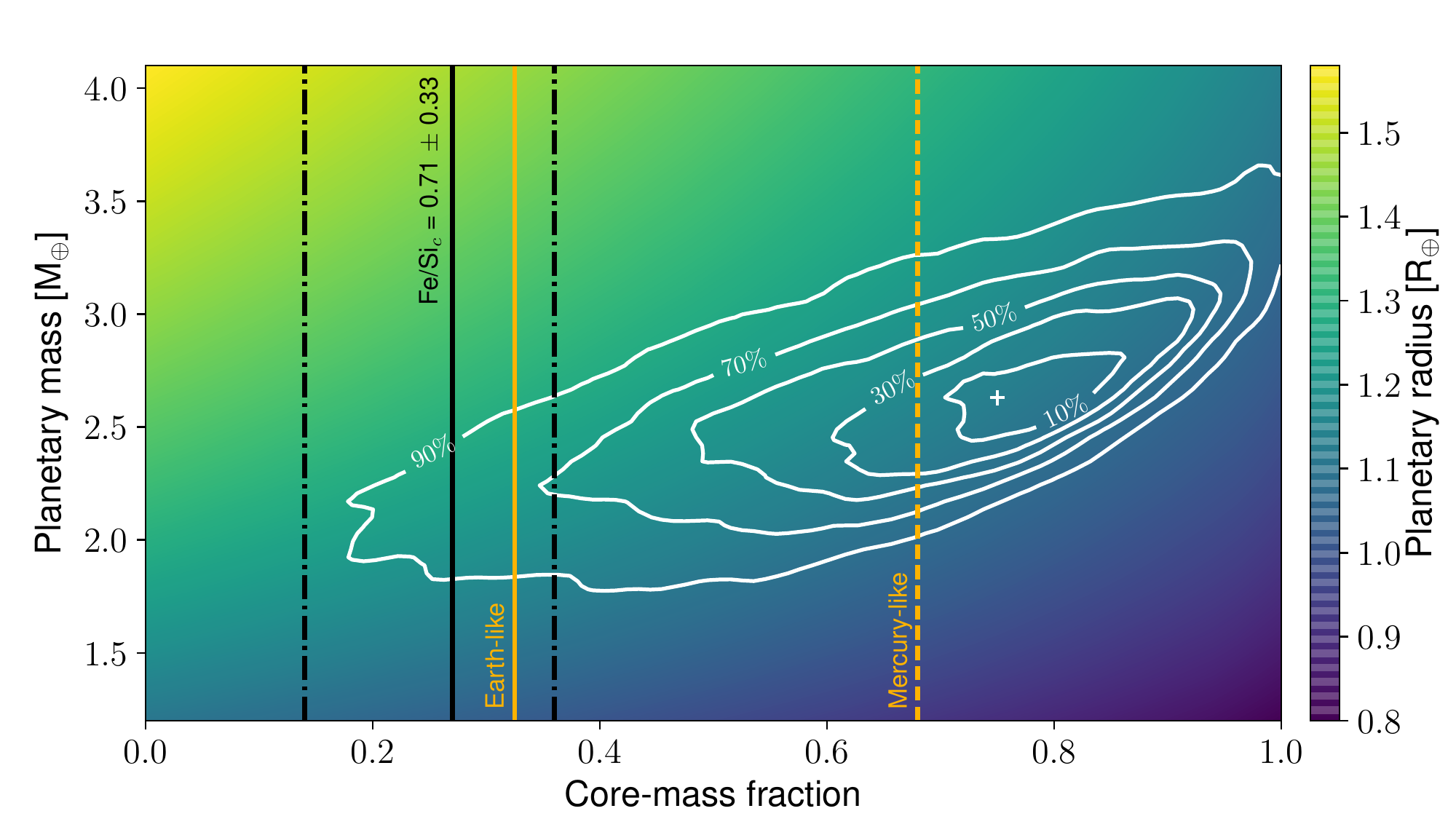}
\label{fig3}
\caption{\textbf{Theoretical radii of dry terrestrial worlds as a function of their total mass and core-mass fraction (assuming no water) for the planet K2-229~b.} The ``+'' mark represents the mode of the posterior distribution. The white contours represent the 10\%, 30\%, 50\%, 70\%, and 90\% percentiles of the posterior distribution. The vertical solid and dot-dashed black lines represent the core-mass fraction and 1-$\sigma$ uncertainties, that corresponds to Fe/Si$_{c}$ = $0.71\pm0.33$ as measured in the host star. The vertical solid and dashed orange lines indicate the Earth- and Mercury-like core-mass fractions, respectively. }
\end{center}
\end{figure}

\begin{table}
\centering
\caption{List of the main physical parameters of the K2-229 planetary system.}
\medskip
\begin{footnotesize}
\begin{tabular}{lccc}
\hline
Parameter & \multicolumn{3}{c}{Value and 68.3\% credible interval}\\
\hline
 & Host star\\
Effective temperature \teff\ [K] & $5185 \pm 32$\\
Surface gravity \logg\ [cgs] & $4.56^{_{+0.03}}_{^{-0.05}}$\\
Iron abundance \met\ [dex] & $-0.06 \pm 0.02$\\
Mass M$_{\star}$\ [\Msun] & $0.837^{_{+0.019}}_{^{-0.025}}$\\
Radius R$_{\star}$\ [\Rsun] & $0.793^{_{+0.032}}_{^{-0.020}}$\\
Age $\tau$\ (stellar evolution tracks) [Gyr] & $5.4^{_{+5.2}}_{^{-3.7}}$\\
Age $\tau$\ (abundance ratio) [Gyr] & $3.0\pm0.9$\\
Distance to Earth D [pc] & $104 \pm 4$\\
Rotation velocity \vsini\ [\kms] & $2.4\pm0.5$\\
Rotation period P$_{\rm rot}$ [d] & $18.1\pm0.3$\\
Activity index logR'$_{\rm H\&K}$ & $-4.58\pm0.04$\\
& & &\\
 & Planet b & Planet c & Planet d\\
Period P [d] & $0.584249 \pm 1.4\times10^{-5}$ & $8.32834 \pm 4.5\times10^{-4}$ & $31.0 \pm 1.1$\\
Eccentricity e & $0$ (assumed) & $0$ (assumed) & $0.39 \pm 0.29$\\
Semi-major axis a [AU] & $0.012888 \pm 1.3\times10^{-4}$ & $0.07577 \pm 7.6\times10^{-4}$ & $0.1820 \pm 4.2\times10^{-3}$\\
Inclination i [\degr] & $83.9 \pm 2.8$ & $87.94 \pm 0.18$ & $88.92 \pm 0.24$\\
%& & &\\
Radius R$_{p}$ [\Rearth] & $1.164^{_{+0.066}}_{^{-0.048}}$ & $2.12^{_{+0.11}}_{^{-0.08}}$ & $2.65 \pm 0.24$\\
Mass M$_{p}$ [\Mearth] & $2.59 \pm 0.43$ & $<21.3^{\dag}$ & $<25.1^{\dag}$\\
Bulk density $\rho_{p}$ [\gcm3] & $8.9 \pm 2.1$ & $<12.8^{\dag}$ &  $<9.5^{\dag}$\\
Equilibrium temperature$^{\ast}$ T$_{eq}$ [K] & $1960 \pm 40$ & $800 \pm 20$ & $522 \pm 13$\\
Day-side temperature$^{\ast\ast}$ T$_{d}$ [K] & $2332 \pm 56$ & $962 \pm 23$ & -- \\
\hline
& & &\\
%\multicolumn{4}{l}{$^{\dag}$rBJD = BJD - 2457000}\\
\multicolumn{4}{l}{$^{\dag}$95\% credible upper limit}\\
\multicolumn{4}{l}{$^{\ast}$Assuming a zero albedo.}\\
\multicolumn{4}{l}{$^{\ast\ast}$Assuming a zero albedo and a tidally synchronised rotation.}\\
\end{tabular}
\end{footnotesize}
\end{table}

\end{document}